\documentclass[reprint,showpacs,amsmath,amssymb,aps,pre]{revtex4-1}
\usepackage{dcolumn}
\usepackage{bm}
\usepackage[dvipdfmx]{graphicx}
\usepackage[dvipdfmx]{hyperref}
\usepackage[mathlines]{lineno}
\usepackage{physics}

\begin{document}

\title{Real-space formulation of topology for disordered Rice-Mele chains without chiral symmetry}
\author{Kiminori Hattori and Ata Yamaguchi}
\affiliation{Department of Systems Innovation, Graduate School of Engineering Science, Osaka University, Toyonaka, Osaka 560-8531, Japan}
\date{\today}

\begin{abstract}	
In this paper, we derive a real-space topological invariant that involves all energy states in the system.
This global invariant, denoted by $Q$, is always quantized to be 0 or 1, independent of symmetries.
In terms of $Q$, we numerically investigate topological properties of the nonchiral Rice-Mele model including random onsite potentials to show that nontrivial bulk topology is sustained for weak enough disorder.
In this regime, a finite spectral gap persists, and then $Q$ is definitely identified.
We also consider sublattice polarization of disorder potentials.
In this case, the energy spectrum retains a gap regardless of disorder strength so that $Q$ is unaffected by disorder.
This implies that bulk topology remains intact as long as the spectral gap opens.
\end{abstract}

\maketitle

\section{Introduction}
\label{sec:1}

Topology residing in quantum states of matter has attracted immense attention in recent years since it produces fundamentally new physical phenomena and potential applications in novel devices.
Historically, the concept of topological matter originates from the theoretical formulation of the integer quantum Hall effect by Thouless et al. \cite{ref:1}.
They showed that the first Chern number, a topological invariant quantized at integers, accounts for the robust quantization of Hall conductivity in two dimensions (2D).
The subsequent discovery of topological insulators in time-reversal-symmetric situations revealed the ubiquitous existence of topological materials, leading to the widespread study of topological aspects in insulators, semimetals, and superconductors \cite{ref:2,ref:3}.
In parallel to these studies, a classification scheme was established that predicts what topological invariants are definable for given systems \cite{ref:4,ref:5}.
This scheme depends only on spatial dimensions and underlying symmetries including time-reversal symmetry, particle-hole symmetry, and the combination of the two known as chiral symmetry.
For instance, the quantum Hall states breaking time-reversal symmetry are classified into the unitary class A, which has a $\mathbb{Z}$ invariant in 2D, while the time-reversal-symmetric topological insulators belong to the simplectic class AII, which allows a ${\mathbb{Z}_2}$ invariant in 2D or 3D.

The well-known Su-Schrieffer-Heeger (SSH) model forms a prototype for topological insulators in 1D, which consists of a bipartite lattice with two sublattice sites in each unit cell \cite{ref:6}.
This model preserves chiral symmetry and belongs to the chiral-orthogonal class BDI, which possesses a $\mathbb{Z}$ invariant known as the winding number in 1D \cite{ref:5}.
The SSH model is generalized to the Rice-Mele (RM) model by including a staggered sublattice potential \cite{ref:7}.
It is well known that adiabatic charge pumping is enabled in the RM model when the system parameters are modulated periodically and slowly with time \cite{ref:8,ref:9,ref:10}.
The charge pump transports topologically quantized charge between neighboring cells without a driving electric field.
The resulting adiabatic current is analogous to the dissipationless quantum Hall current.
As demonstrated by Thouless \cite{ref:11}, the pumped charge per cycle is explicitly described by the Berry curvature in time-momentum space and the associated Chern number.
The charge pumping also provides a firm foundation for the modern theory of electric polarization, which is connected to the Berry phase across the Brillouin zone \cite{ref:8,ref:9,ref:10}.
The topological charge pump has been realized in recent experiments utilizing ultracold atoms \cite{ref:12,ref:13}.
Theoretically, it is also shown that energy pumping is feasible in this system \cite{ref:14}.
	
The onsite potentials violate chiral symmetry so that the RM model falls in the orthogonal class AI, which allows no topological index in 1D.
Hence, this model is not a topological insulator in terms of the prevailing classification scheme.
This argument is basically valid insofar as the subspace of occupied (or unoccupied) states is concerned.
Removing the restriction to take all energy states into consideration alters the conclusion.
In the previous studies \cite{ref:14,ref:15,ref:16}, such a global invariant is formulated in momentum space for the RM model.
The formulation is however incomplete, particularly in the sense that it is inapplicable to disordered systems lacking translational invariance.
In this manuscript, we generalize the theoretical treatment to derive a real-space formula, which can deal with ordered and disordered systems on an equal footing.

The paper is organized as follows.
To be self-contained, we summarize the momentum-space formulation for ordered nonchiral RM chains in Sec. \ref{sec:2}.
In Sec. \ref{sec:3}, we define a real-space topological invariant that involves all energy states in the system.
This global invariant, denoted by $Q$ in the following, is always quantized to be 0 or 1, independent of symmetries.
We also show that the two formulations are equivalent in the absence of disorder.
In terms of $Q$, we numerically elucidate topological properties of the systems subjected to random onsite potentials in Sec. \ref{sec:4}.
It is shown from numerical results that nontrivial bulk topology is sustained for weak enough disorder.
In this regime, a finite spectral gap remains, and then $Q$ is definitely identified.
This argument is supported by the persistence of topological edge modes in the gapped regime.
We also introduce sublattice polarization into disorder potentials.
In this case, the finite gap persists regardless of disorder strength so that $Q$ is unaffected by disorder.
This observation implies that bulk topology remains intact as long as the spectral gap is open.
Finally, Sec. \ref{sec:5} provides a summary.

\section{Model and winding number}
\label{sec:2}

The RM model is described by the Hamiltonian $H = {\sum _{jj'\alpha \beta }} \ket{j,\alpha } H_{jj'}^{\alpha \beta } \bra{j',\beta } $ in real space for noninteracting spinless fermions, where $j \in \{ 1,2, \cdots ,N\} $ denotes the lattice position of the two-site unit cell, $\alpha ,\beta \in \{ A,B\} $ represents the sublattice degree of freedom in each cell, and $\ket{j,\alpha } $ is the basis ket at each site.
The matrix element $H_{jj'}^{\alpha \beta }$ is explicitly written as $H_{jj'}^{AB} = H_{j'j}^{BA} = v{\delta _{jj'}} + w{\delta _{j,j' + 1}}$ and $H_{jj'}^{AA} = - H_{jj'}^{BB} = m{\delta _{jj'}}$, where $v$ ($w$) denotes the intracell (intercell) hopping energy, and $m$ describes the staggered sublattice potential.
For simplicity, we assume that $v$ and $w$ are nonnegative in the following.
Note that if $m = 0$, the RM model is reduced to the SSH model with chiral symmetry.
In momentum space, the Hamiltonian is formulated as $H(k) = {\mathbf{h}}(k) \cdot \boldsymbol{\sigma} $, where $\boldsymbol{\sigma} = ({\sigma _x},{\sigma _y},{\sigma _z})$ is the Pauli vector, and the 3D vector ${\mathbf{h}} = ({h_x},{h_y},{h_z})$ is composed of ${h_x} = v + w \cos 2k$, ${h_y} = w \sin 2k$, and ${h_z} = m$.
In the present formulation, we assume that lattice sites are equally spaced, and take the intersite distance as the unit of length.
In this unit, the intercell distance is 2.

The eigenequation $H(k) \ket{{u_r}(k)} = {\varepsilon _r}(k) \ket{{u_r}(k)}$ is solved to be ${\varepsilon _r} = rh$ and
\begin{equation*}
\ket{u_r} = \frac{1}{\sqrt {2h(h - r{h_z})}} \left( {\begin{array}{c} {r({h_x} - i{h_y})} \\ {h - r{h_z}} \end{array}} \right) ,
\end{equation*}
where $r = \pm 1$ denotes the band index, and $h = \sqrt {h_x^2 + h_y^2 + h_z^2} $.
The corresponding Berry connection is given by
\begin{equation}
\label{eq:1}
{A_r} = i \mel{u_r}{\frac{\partial }{\partial k}}{u_r} = \frac{{{h_x}\frac{{\partial {h_y}}}{{\partial k}} - {h_y}\frac{{\partial {h_x}}}{{\partial k}}}}{{2h(h - r{h_z})}} ,
\end{equation}
for band $r$.
In terms of ${A_r}$, the Berry phase across the 1D Brillouin zone $k \in (0,\pi ]$, i.e., the Zak phase, is expressed as ${\nu _r} = {(2\pi )^{ - 1}}\smallint _0^\pi dk{A_r}$ in units of $2\pi $.
This quantity is not quantized unless $m = 0$, in agreement with the topological classification in terms of symmetries and dimensions.
However, summing over two bands, we obtain
\begin{equation}
\label{eq:2}
\sum\limits_r {{A_r}} = \frac{{{h_x}\frac{{\partial {h_y}}}{{\partial k}} - {h_y}\frac{{\partial {h_x}}}{{\partial k}}}}{{h_x^2 + h_y^2}} = \frac{{\partial \phi }}{{\partial k}} ,
\end{equation}
where $\phi = {\tan ^{ - 1}}({h_y}/{h_x})$.
Hence, the total Berry phase,
\begin{equation}
\label{eq:3}
\nu = \sum\limits_r {{\nu _r}} = \frac{1}{{2\pi }}\int_0^\pi {dk\frac{{\partial \phi }}{{\partial k}}} ,
\end{equation}
amounts to a winding number that describes how often the 2D vector $({h_x},{h_y})$ winds about the origin as $k$ varies over the entire Brillouin zone \cite{ref:14,ref:16}.
The winding number $\nu $ is a topological invariant and is generically integer quantized.
Notably, $\nu $ is independent of $m$.
The same conclusion is derived also from the relative phase between components of the eigenvector \cite{ref:15,ref:17}. 
The SSH model preserves chiral symmetry.
In this case, ${\nu _r} = \nu /2$ is quantized equally for two bands at a half-integer.
Note that a particular gauge is chosen in the above argument.
To derive a gauge-invariant expression, the momentum-space formulation will be revisited in a different manner in Sec. \ref{sec:3}.

\begin{figure}
\centering
\includegraphics[width=\linewidth]{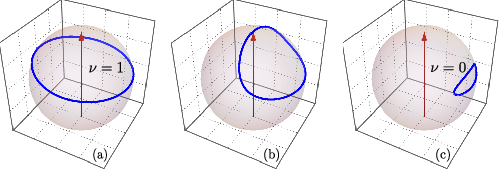}
\caption{(Color online) A closed loop formed on the Bloch sphere by the 3D unit vector ${\mathbf{\hat h}} = {\mathbf{h}}/h$ as $k$ varies across the Brillouin zone. In each panel, the arrow represents the $z$ axis. The hopping energies are varied as $(v,w) = (1,2)$ for (a), $(1.5,1.5)$ for (b), and $(2,1)$ for (c). These results are derived for $m = 0.5$.}
\label{fig:1}
\end{figure}

A geometrical meaning of $\nu$ is clarified in Fig. \ref{fig:1}.
The 3D unit vector ${\mathbf{\hat h}} = {\mathbf{h}}/h$ forms a closed loop on the Bloch sphere as $k$ goes across the Brillouin zone.
The topological invariant $\nu$ accounts for the number of times ${\mathbf{\hat h}}$ (or $\mathbf{h}$) passes around the $z$ axis.
Thus, $\nu = 1$ for $v < w$ and 0 for $v > w$.
Note that the closed curve is deformed by varying the parameters $(v,w)$ and intersects the $z$ axis at $v=w$, when the topological phase transition occurs.

The nontrivial topology of the RM model is supported by the bulk-boundary correspondence \cite{ref:14,ref:17}.
In a finite RM chain with open boundaries, two ingap edge modes emerge in the nontrivial phase $v < w$.
Thus, the presence or absence of edge modes correlates to the bulk topology defined by the winding number $\nu $.
In the limit of $N \to \infty $, the eigenfunctions are analytically derived for the two edge modes to be $\ket{L} = c{\sum _j} \ket{j,A} {( - v/w)^{j - 1}}$ and $\ket{R} = c{\sum _j} \ket{j,B} {( - v/w)^{N - j}}$, where $c = \sqrt {1 - v^2 / w^2} $ is the normalization constant.
These solutions fulfill the eigenequations $H \ket{L} = m \ket{L} $ and $H \ket{R} = - m \ket{R} $.
The existence condition of edge modes is $v < w$.
This criterion is identical to that for $\nu = 1$.
Note that $\ket{L}$ and $\ket{R}$ are independent of $m$ and valid for both RM and SSH models.

\section{Real-space topological invariant}
\label{sec:3}

In the presence of the inevitable disorder in real systems, the momentum-space formulation is no longer verifiable.
In what follows, we derive the real-space topological invariant in analogy to Resta polarization \cite{ref:18} and its extension to multipole moments in higher-order topological insulators \cite{ref:19,ref:20,ref:21,ref:22,ref:23,ref:24,ref:25,ref:26}.
To this end, it is convenient to introduce a simpler representation such that $H = {\sum _{\left| {l - l'} \right| < 2}} \ket{l} H_{ll'} \bra{l'} $, where $l \in \{ 1,2, \cdots ,2N\} $ denotes the site index.
For the RM model, the matrix element is given by ${H_{l,l+1}} = {H_{l+1,l}} = v$ for $l \in {\text{odd}}$ and $w$ for $l \in {\text{even}}$, and ${H_{ll}} = m$ for $l \in {\text{odd}}$ and $ - m$ for $l \in {\text{even}}$.
Diagonal onsite disorder is described by $V = {\sum _l} \ket{l} {V_l} \bra{l}$.
The total Hamiltonian consists of $\mathcal{H} = H + V$.
Note that chiral symmetry is broken by including $V$ as well as the mass term $m$ in $H$.
The eigenequation is expressed as $\mathcal{H} \ket{u_{rn}} = {\varepsilon _{rn}} \ket{u_{rn}} $, where $n \in \{ 1,2, \cdots ,N\} $ and $r = \pm 1$.
The positive and negative signs correspond to the upper and lower halves of the eigenenergies, respectively.
The rectangular matrix $({U_r})_{ln} = \braket{l}{u_{rn}} $ constructed from the eigenvectors in $r$ subspace obeys $U_r^\dag {U_r} = I$ and ${\sum _r}{U_r}U_r^\dag = I$, where $I$ denotes the identity matrix.
The position matrix ${X_{ll'}} = {x_l}{\delta _{ll'}}$ defines the momentum translation matrix as $T = {e^{i \delta kX}}$, where $\delta k = \pi /N$ is the momentum interval in a periodic chain of size $2N$.
Then, Sylvester's determinant identity $\det (I+AB) = \det (I+BA)$ leads to $\det F_r = \det F_{ - r}^\dag \det T$, where ${F_r} = U_r^\dag T{U_r}$ \cite{ref:22}.
This relation ensures that ${z_r} = z_{ - r}^*$ for the complex number ${z_r} = \det F_r \sqrt {\det T^\dag } $.
Thus, the real-space topological index is simply definable as
\begin{equation}
\label{eq:4}
Q = \frac{1}{\pi } \arg \sum\limits_r {{z_r}} ,
\end{equation}
where $\arg z \in ( - \pi ,\pi ]$ is the principal value of the argument.
Note that $Q$ is a global index that involves all energy states.
Since ${\sum _r}{z_r}$ is real, $Q$ is quantized to be 0 or 1.
Importantly, the ${\mathbb{Z}_2}$-quantization is validated regardless of symmetries.
It is also worth noting that $Q$ is explicitly gauge invariant.
Because of ${\sum _r}{z_r} = 2 {\Re} {z_ + } = 2 {\Re} {z_ - }$, $Q$ can be reformulated in a different form ${(-1)^Q} = \operatorname{sgn} {\Re} {z_ + } = \operatorname{sgn} {\Re} {z_ - }$.
Note that we can freely choose the origin of the coordinate to measure site positions $\{ {x_l}\} $.
Symmetrizing $\{ {x_l}\} $ so that $\Tr X = 0$, the complex number is reduced to ${z_r} = \det F_r $.

As mentioned above, no specific symmetries are assumed for deriving $Q$.
If the system respects chiral symmetry, the eigenstates in two subspaces are related as ${\sigma _z}{U_r} = {U_{ - r}}$.
In this case, $\det F_r = \det F_{ - r} $, and thereby ${z_r} = z_r^*$ is real.
Thus,
\begin{equation}
\label{eq:5}
{Q_r} = \frac{1}{{2\pi }} \arg {z_r} ,
\end{equation}
is quantized to be 0 or 1/2.
It is also shown from ${z_r} \in \mathbb{R}$ that $Q = 2{Q_ + } = 2{Q_ - }$.
It should be emphasized that for nonchiral systems, ${z_r} \notin \mathbb{R}$ and hence ${Q_r}$ is no longer quantized.
Thus, the present argument does not contradict the prevailing theory for symmetry-protected topological insulators.

A remaining question is the relation between the momentum-space invariant $\nu $ and the real-space invariant $Q$ in the absence of disorder.
For the ordered periodic RM model, the real-space eigenequation is written as ${\sum _{j'\beta }}H_{jj'}^{\alpha \beta }{b_{rk\beta }}(j') = {\varepsilon _r}(k){b_{rk\alpha }}(j)$, where ${b_{rk\alpha }}(j) = {N^{ - 1/2}}\braket{\alpha}{u_r (k)} {e^{ik{x_j}}}$ is the Bloch function.
It is convenient to symmetrize cell positions such that ${x_j} = 2(j - \tfrac{{N + 1}}{2})$.
Then, site positions consist of ${x_l} \in \{ {x_{1A}},{x_{1B}}, \cdots ,{x_{NA}},{x_{NB}}\} $, where ${x_{jA}} = {x_j} - 1/2$ and ${x_{jB}} = {x_j} + 1/2$.
In this definition, ${\sum _l}{x_l} = 0$ so that ${z_r} = \det F_r$ as mentioned before.
Note that the matrix ${F_r}$ is generically expressed as ${({F_r})_{nn'}} = {\sum _l}u_{rn}^*(l){e^{i\delta k{x_l}}}{u_{rn'}}(l)$ in terms of the real-space eigenfunction ${u_{rn}}(l) = \braket{l}{u_{rn}}$.
Thus, we have ${({F_r})_{kk'}} = {\sum _{j\alpha }}b_{rk\alpha }^*(j){e^{i\delta k{x_{j\alpha }}}}{b_{rk'\alpha }}(j)$ for the Bloch eigenstate specified by the discrete wavenumber $k$.
It is easy to show that this formula leads to ${({F_r})_{kk'}} = \braket{{{\tilde u}_r}(k)}{{{\tilde u}_r}(k')} {\delta _{k',k - \delta k}}$, where $\ket{{{\tilde u}_r}(k)} = {e^{i k{\sigma _z}/2}} \ket{{u_r}(k)} $ is the unitary transformed eigenvector.
The determinant $\det F_r = {\prod _k} \braket{{{\tilde u}_r}(k)}{{{\tilde u}_r}(k - \delta k)} $ forms a Wilson loop across the Brillouin zone so that
\begin{equation}
\label{eq:6}
{\tilde \nu _r} = \frac{1}{{2\pi i}} \ln {z_r} = \frac{1}{{2\pi }}\int_0^\pi {dk{{\tilde A}_r}} ,
\end{equation}
in the limit of $N \to \infty $, where ${\tilde A_r} = i \mel{{\tilde u}_r}{\partial _k}{{\tilde u}_r} $ is the relevant Berry connection.
This result amounts to
\begin{equation}
\label{eq:7}
Q = \frac{1}{\pi } \arg \sum\limits_r {e^{2\pi i {{\tilde \nu }_r}}} ,
\end{equation}
\begin{equation}
\label{eq:8}
{Q_r} = \frac{1}{{2\pi }}\arg {e^{2\pi i {{\tilde \nu }_r}}} .
\end{equation}
The Berry connection consists of ${\tilde A_r} = {A_r} - \tfrac{1}{2} \mel{u_r}{\sigma _z}{u_r} $.
Hence, we obtain ${\tilde \nu _r} = {\nu _r} - r \eta $, where $\eta = {(4\pi )^{ - 1}} \smallint _0^\pi dk{h_z}/h$.
Assuming chiral symmetry, ${h_z} = \eta = 0$ so that ${\tilde \nu _r} = {\nu _r}$.
This means ${Q_r} = {\nu _r}$.
It is also shown from ${\nu _r} = \nu /2$ that $Q = \nu $.
A simple equation ${\sum _r}{e^{2\pi i {\tilde \nu }_r}} = 2{e^{i\pi \nu }} \cos \theta $ is useful in dealing with nonchiral systems, where $\theta = \tfrac{1}{2}\smallint _0^\pi dk({\partial _k}\phi - 1){h_z}/h$.
It is easy to show that $\left| \theta \right| < \left| {\theta _0} \right| = \tfrac{\pi }{2}$ and hence $\cos \theta > 0$, where ${\theta _0} = \tfrac{1}{2}\smallint _0^\pi dk({\partial _k}\phi - 1)$.
Therefore, $Q$ and $\nu $ are generically identical in the absence of disorder.
For simplicity, a fixed gauge is used in the above calculation.
However, the gauge choice is irrelevant to the argument, since $Q$ and ${Q_r}$ are gauge invariant.

\begin{figure}
\centering
\includegraphics[width=\linewidth]{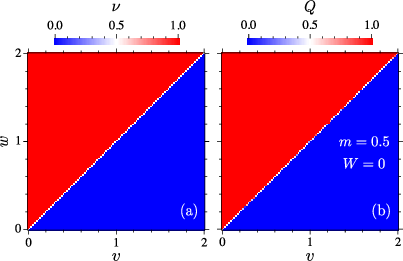}
\caption{(Color online) Phase diagrams of (a) momentum-space winding number $\nu $ and (b) real-space topological invariant $Q$ in parameter space $(v,w)$ for ordered chains without disorder ($W = 0$). The latter is derived for $m = 0.5$ and $N = 500$.}
\label{fig:2}
\end{figure}

Figure \ref{fig:2} compares $\nu $ and $Q$.
The real-space invariant $Q$ is numerically derived for a periodic ordered chain with $m = 0.5$ and $N = 500$.
As seen in the figure, two invariants are identically quantized to be 0 or 1, depending on relative magnitudes of $v$ and $w$.
At $v = w$, the topological phase transition takes place.
Thus, the bulk topology of nonchiral RM chains is clearly identified in terms of $Q$.

\section{Numerical calculations}
\label{sec:4}

Next, we proceed to numerical calculations for disordered systems.
Considering sublattice degrees of freedom, random onsite potentials are expressed generally as $V = {\sum _{j\alpha }} \ket{j,\alpha } V_j^\alpha \bra{j,\alpha }$.
We regard $V_j^\alpha \in [ - {W_\alpha },{W_\alpha }]$ as a random variable following the uniform distribution.
For normal disorder, we choose $W = {W_A} = {W_B}$.
On the other hand, either ${W_A} = 0$ or ${W_B} = 0$ is assumed for polarized disorder.
Off-diagonal bond disorder, i.e., disorder in hopping energies $v$ and $w$, preserves chiral symmetry of the SSH model.
We neglect such a special class of disorder in this study.
In numerical calculations, disorder averaging is performed over ${10^3}$-${10^4}$ random configurations unless stated otherwise.
The mean and standard deviation of a quantity $O$ are denoted by $\ev{O}$ and $\Delta O$, respectively.
The parameters used in the calculation are typically $m = 0.5$ and $N = 500$.

\subsection{Normal disorder}
\label{sec:4A}

\begin{figure}
\centering
\includegraphics[width=\linewidth]{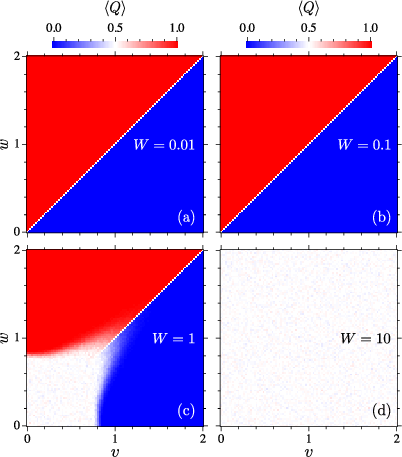}
\caption{(Color online) Maps of disorder-averaged topological invariant $\ev{Q}$ in parameter space $(v,w)$. The disorder strength $W$ is varied as (a) 0.01, (b) 0.1, (c) 1, and (d) 10. Note that the panel (d) shows $\ev{Q} \simeq 0.5$ in the entire parameter range. These results are derived for $m = 0.5$ and $N = 500$.}
\label{fig:3}
\end{figure}

\begin{figure}
\centering
\includegraphics[width=\linewidth]{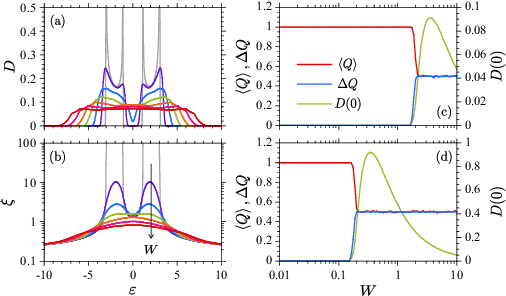}
\caption{(Color online) The left two panels show (a) density of states $D(\varepsilon )$ and (b) localization length $\xi (\varepsilon )$ for $(v,w,m) = (1,2,0.5)$ in the $N \to \infty $ limit. In each panel, $W$ is varied as 0.1, 1, 2, 3, 4, 5 and 6. In the two panels on the right hand side, $D(0)$ is compared to $\ev{Q}$ and $\Delta Q$ for $N = 500$ at various $W$'s. The parameters used in these calculations are $(v,w,m) = (1,2,0.5)$ for (c) and $(0.1,0.2,0)$ for (d).}
\label{fig:5}
\end{figure}

In this subsection, we discuss the numerical results for normal disorder with $W = {W_A} = {W_B}$.
Figure \ref{fig:3} shows how $\ev{Q}$ varies in $(v,w)$ space at various $W$'s.
As seen in the figure, $\ev{Q}$ remains quantized to be 0 or 1 for weak disorder ($W \ll 1$).
In this case, the phase boundary separating trivial and nontrivial phases is clearly visible at $v = w$.
For medium disorder ($W \approx 1$), $\ev{Q} = 0.5$ is observed in a specific region where $v$ and $w$ are relatively small.
The corresponding area covers the entire parameter range for strong disorder ($W \gg 1$).
In the regions where $\ev{Q} = $0 or 1, $\Delta Q$ vanishes.
On the other hand, $\Delta Q = 0.5$ for $\ev{Q} = 0.5$ (see, Fig. \ref{fig:5}).
The latter implies that $Q$ takes two possible integers with equal probabilities for each single realization of disorder.
In other words, $Q$ is indefinite in this regime.
This behavior is reminiscent of that in the Griffiths phase where trivial and nontrivial samples coexist \cite{ref:23,ref:27}.
As shown later, this regime relates to closure of the bulk spectral gap.

\begin{figure}
\centering
\includegraphics[width=\linewidth]{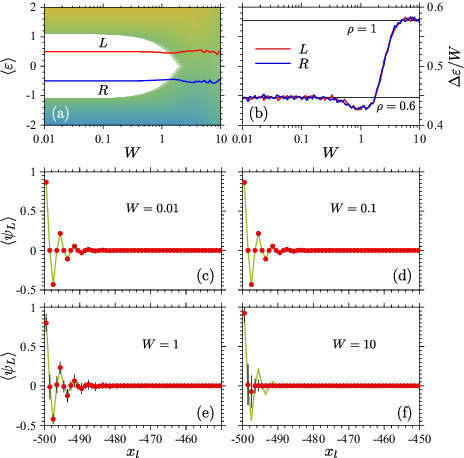}
\caption{(Color online) The upper two panels show (a) disorder-averaged eigenenergies $\ev{\varepsilon } $ for bulk and edge modes, and (b) normalized standard deviations $\Delta \varepsilon /W$ for edge modes, as a function of $W$. The two thin lines in (b) indicate the analytic results for $\rho = 0.6$ and 1. The lower four panels display disorder-averaged wavefunctions $\ev{\psi _L}$ of the edge mode localized at the left end. The disorder strength $W$ is varied as (c) 0.01, (d) 0.1, (e) 1, and (f) 10. In each panel, a solid line shows the analytic wavefunction for $W = 0$, and dots and error bars denote means and standard deviations, respectively. All of these results are derived for $(v,w,m) = (1,2,0.5)$ and $N = 500$.}
\label{fig:4}
\end{figure}

Figure \ref{fig:4} (a) shows the disorder-averaged eigenvalues for finite chains with $v = 1$ and $w = 2$ as a function of $W$.
In the calculation, each numerical eigenstate $\ket{u_{rn}} $ is projected onto the analytic ones $\ket{L} $ and $\ket{R} $ for ordered semi-infinite chains.
The edge mode is identified as the eigenstate that maximizes the relevant overlap $\left| \braket{L}{u_{rn}} \right|$ or $\left| \braket{R}{u_{rn}} \right|$ \cite{ref:28}.
As seen in the figure, the averaged eigenvalues of these edge modes do not largely deviate from $ \pm m$.
Analogous results are reported in the literature for the SSH model including onsite disorder \cite{ref:28,ref:29,ref:30}.
Basically, this behavior is accounted for in the framework of perturbation theory, which predicts ${\varepsilon _L} = m + {V_L}$ and ${\varepsilon _R} = - m + {V_R}$ to first order, where ${V_L} = \mel{L}{V}{L} $ and ${V_R} = \mel{R}{V}{R} $.
Since $\ev{V_L} = \ev{V_R} = 0$ for the assumed random potentials, it is easy to see $\ev{\varepsilon _L} = m$ and $\ev{\varepsilon _R} = - m$.
It is also noticed that $\ev{V_L^2} = \rho W_A^2/3$ and $\ev{V_R^2} = \rho W_B^2/3$, where $\rho = ({w^2} - {v^2})/({w^2} + {v^2})$ is the inverse participation ratio for $\ket{L}$ and $\ket{R}$.
Hence, one expects $\Delta {\varepsilon _L} = {W_A}\sqrt {\rho /3} $ and $\Delta {\varepsilon _R} = {W_B}\sqrt {\rho /3} $.
For weak disorder, the analytic results derived with $\rho = 0.6$ for the assumed parameters agree with the numerical ones, as shown in Fig. \ref{fig:4} (b).
For strong disorder, the numerical results are more reasonably explained by $\rho = 1$.
This implies that the relevant edge modes are Anderson localized due to strong disorder.
The lower panels (c)-(f) of Fig. \ref{fig:4} summarize the disorder-averaged wavefunctions $\ev{\psi _L}$ at various $W$'s for the edge mode localized at the left end.
For weak disorder, the numerical results are identical to the analytic wavefunction, exemplifying the persistence of a topological edge mode.
The suggested Anderson localization is clearly seen for strong disorder.
For this reason, the relevant edge states are indistinguishable from ordinary localized states with no topological origin.
The observation is similar for the edge mode localized at the opposite boundary.
The eigenstates other than edge states constitute two dense bands in the disorder-averaged energy spectrum.
The spectral gap separating two bands decreases as $W$ increases, and vanishes for $W \geq 2$.

The density of bulk states and their localization lengths are evaluated by using the retarded Green's function formulated in a recursive manner \cite{ref:31,ref:32,ref:33,ref:34} as
\begin{equation}
\label{eq:9}
{G_{MM}} = {(g_M^{-1} - {\mathcal{H}_{M,M-1}}{G_{M-1,M-1}}{\mathcal{H}_{M-1,M}})^{-1}} ,
\end{equation}
\begin{equation}
\label{eq:10}
{G_{1M}} = {G_{1,M-1}}{\mathcal{H}_{M-1,M}}{G_{MM}} ,
\end{equation}
for an open chain consisting of $M$ sites, where ${g_M} = {(\varepsilon - {\mathcal{H}_{MM}})^{-1}}$ represents the Green's function of an isolated site.
Combining the diagonal element ${G_{MM}}$ with the two auxiliary functions ${J_M} = (1+{\mathcal{H}_{M,M-1}}{K_{M-1}}{\mathcal{H}_{M-1,M}}){G_{MM}}$ and ${K_M} = {J_M}{G_{MM}}$, the density of states per site is derived to be
\begin{equation}
\label{eq:11}
D = - \lim \limits_{N \to \infty } \frac{1}{{2\pi N}}\sum\limits_{M=1}^{2N} {\Im J_M} ,
\end{equation}
in the thermodynamic limit.
Note that two edge states are irrelevant to $D$ in comparison to $2N - 2$ bulk states in the $N \to \infty $ limit.
In terms of the off-diagonal element ${G_{1M}}$ joining two ends, the localization length is formulated as
\begin{equation}
\label{eq:12}
\xi = - \lim \limits_{N \to \infty } \frac{{N-1}}{{\ln \left| {{G_{1,2N}}/{G_{12}}} \right|}} ,
\end{equation}
per cell.
By definition, $\xi $ is dominated by the energy states most extended along the chain.
Since the edge states are localized even in the absence of disorder, their contributions to $\xi $ are normally minor as shown later.

Figure \ref{fig:5} summarizes the numerical results for various $W$'s.
As shown in Fig. \ref{fig:5} (a), $D$ is composed of upper and lower bands separated by a finite gap for weak disorder.
As $W$ increases, these two bands appreciably broaden and eventually merge.
The resulting gap closure is observed for $W \geq 2$.
It is shown in Fig. \ref{fig:5} (b) that $\xi $ is maximal at the center of each band for weak disorder.
For strong disorder, $\xi $ reduces sizably, and its spectrum deforms into a single band peaked at $\varepsilon = 0$.
This indicates that the gap is filled with strongly localized states.
Recall that edge states exhibit an exponential decay in an ordered chain.
The decay length is given by ${\xi _{{\text{edge}}}} = 1/ \ln (w/v) = 1.44$ for the assumed parameters.
For strong enough disorder, we see $\xi < {\xi _{{\text{edge}}}}$ in the entire energy range, implying Anderson localization of edge states as well as bulk states.

In Figs. \ref{fig:5} (c) and (d), the density of states $D(0)$ at $\varepsilon = 0$ is compared to the disorder-averaged topological invariant $\ev{Q}$ and its standard deviation $\Delta Q$ at various $W$'s.
As shown in these figures, $\ev{Q} = 1$ and $\Delta Q = 0$ as long as $D(0)$ vanishes.
For strong disorder, the spectral gap closes and then $D(0)$ becomes finite.
In this regime, we observe $\ev{Q} = \Delta Q = 0.5$.
These results indicate that bulk topology is sustained for weak enough disorder.
In this regime, two bands are separated by a finite gap, and then $Q$ is definitely identified.
In the gapless regime, $Q$ is indefinite so that bulk topology is no longer identifiable in terms of $Q$.

\subsection{Polarized disorder}
\label{sec:4B}

Recall that ${\varepsilon _L} = m + \mel{L}{V}{L} $ and ${\varepsilon _R} = - m + \mel{R}{V}{R} $ follow from perturbation theory.
In terms of the sublattice polarization of $\ket{L} $ and $\ket{R} $, it is easy to see that $\mel{L}{V}{L} = 0$ for ${W_A} = 0$, and similarly $\mel{R}{V}{R} = 0$ for ${W_B} = 0$.
This suggests that one of the topological edge modes is preserved in the presence of polarized disorder.

\begin{figure}
\centering
\includegraphics[width=\linewidth]{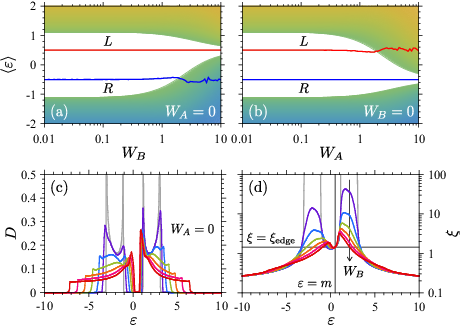}
\caption{(Color online) The upper two panels show disorder-averaged eigenenergies $\ev{\varepsilon } $ versus (a) ${W_B}$ for ${W_A} = 0$ and (b) ${W_A}$ for ${W_B} = 0$. These results are derived for $(v,w,m) = (1,2,0.5)$ and $N = 500$. The lower two panels display (c) density of states $D(\varepsilon )$ and (d) localization length $\xi (\varepsilon )$ for ${W_A} = 0$ in the $N \to \infty $ limit. In each panel, ${W_B}$ is varied as 0.1, 1, 2, 3, 4, 5 and 6. The crossed thin lines in (d) indicate the fixed point at $(\varepsilon ,\xi ) = (m,{\xi _{{\text{edge}}}})$. The parameters $(v,w,m)$ used in these calculations are identical to those for (a).}
\label{fig:6}
\end{figure}

\begin{figure}
\centering
\includegraphics[width=\linewidth]{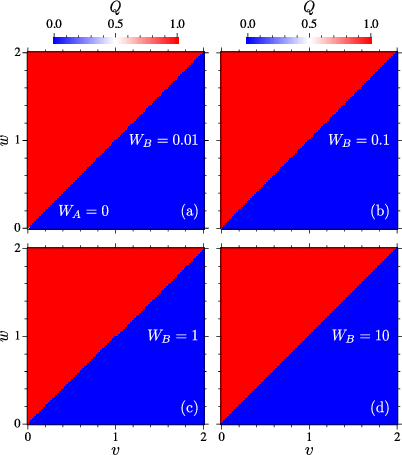}
\caption{(Color online) Phase diagrams of $Q$ in parameter space $(v,w)$ for ${W_A} = 0$. The disorder strength ${W_B}$ is varied as (a) 0.01, (b) 0.1, (c) 1, and (d) 10. The result shown in each panel is derived with a single disorder realization. The parameters used in these calculations are $m = 0.5$ and $N = 500$.}
\label{fig:7}
\end{figure}

Figure \ref{fig:6} displays the numerical results for polarized disorder.
As shown in Fig. \ref{fig:6} (a), $\ev{\varepsilon _L} $ coincides exactly with $m$ irrespective of ${W_B}$ for ${W_A} = 0$.
Analogously, $\ev{\varepsilon _R} = - m$ regardless of ${W_A}$ for ${W_B} = 0$, as seen in Fig. \ref{fig:6} (b).
We have also confirmed that $\Delta {\varepsilon _L} = \Delta {\varepsilon _R} = 0$ for these robust edge modes, and their wavefunctions are identical to those in an ordered chain.
 Although these results agree with perturbation theory, the robustness against strong disorder is not obvious from this perspective.
 Except for the stability of edge modes, an interesting observation for polarized disorder is a finite spectral gap that persists even for strong disorder.
 The gap is located around $\varepsilon = \pm m$ depending on how disorder polarizes.
 The persistence of the spectral gap centered at $\varepsilon = m$ is also confirmed from $D$ for ${W_A} = 0$ shown in Fig. \ref{fig:6} (c).
 Figure \ref{fig:6} (d) displays $\xi $ derived for ${W_A} = 0$.
 As ${W_B}$ increases, $\xi $ decreases in a wide energy range except at $\varepsilon = m$, where $\xi $ remains unchanged.
At this fixed point, we find $\xi = {\xi _{{\text{edge}}}}$.
Thus, this feature stems from the edge mode $\ket{L} $ persisting against polarized disorder with ${W_A} = 0$.

The reason why the gap persists for polarized disorder is accounted for in terms of the self-consistent Born approximation \cite{ref:22,ref:23,ref:33}.
In this approximation, the retarded self-energy matrix is diagonalized as ${\Sigma _{\alpha \beta }} = {\Sigma _\alpha }{\delta _{\alpha \beta }}$ for onsite disorder, and the diagonal elements satisfy the self-consistent equations
\begin{equation}
\label{eq:13}
{\Sigma _A} = \frac{{W_A^2}}{{3N}}\sum\limits_k {\frac{{\varepsilon + m - {\Sigma _B}}}{{\Delta (k)}}} ,
\end{equation}
\begin{equation}
\label{eq:14}
{\Sigma _B} = \frac{{W_B^2}}{{3N}}\sum\limits_k {\frac{{\varepsilon - m - {\Sigma _A}}}{{\Delta (k)}}} ,
\end{equation}
where $\Delta (k) = \det [\varepsilon I - H(k) - \Sigma ]$.
As is easily found, ${\Sigma _A} = 0$ and
\begin{equation*}
{\Sigma _B} = \frac{{W_B^2}}{{3N}}\sum\limits_k {\frac{{\varepsilon - m}}{{\Delta (k)}}} ,
\end{equation*}
for ${W_A} = 0$.
Hence, $\Sigma $ vanishes at $\varepsilon = m$ in this case.
This indicates that the spectral gap survives at this special point even in the presence of disorder.
An analogous result is derived for ${W_B} = 0$.
In this case, $\Sigma = 0$ at $\varepsilon = -m$.
The persisting gap also protects an ingap edge mode by evading hybridization with bulk states.

As shown above, the spectral gap remains open for polarized disorder.
A natural question arising from this peculiarity is how disorder of this type affects the topological invariant $Q$ of our interest.
Figure \ref{fig:7} summarizes $Q$ derived for a single realization of polarized disorder with ${W_A} = 0$.
As seen in the figure, $Q$ is totally unaffected by disorder.
This observation implies that bulk topology remains intact as long as there exists a finite spectral gap.

\section{Summary}
\label{sec:5}

We have derived a real-space topological invariant that involves all energy states in the 1D system.
This global invariant, denoted by $Q$, is always quantized to be 0 or 1, independent of symmetries.
In the absence of disorder, $Q$ is equivalent to the winding number $\nu $ defined in momentum space.
In terms of $Q$, we numerically elucidate topological properties of the nonchiral RM model including random onsite potentials.
It is shown from numerical results that nontrivial bulk topology is sustained for weak enough disorder.
In this regime, a finite spectral gap remains, and then $Q$ is definitely identified.
This argument is corroborated by topological edge modes persisting in the gapped regime.
We also consider sublattice polarization of disorder potentials.
In this case, the energy spectrum stays gapped regardless of disorder strength so that $Q$ is unaffected by disorder.
This observation implies that bulk topology remains intact as long as the spectral gap opens.

\bibliography{ref}
 
\end{document}